# Instability of vortex-antivortex interface in optimally-doped Ba(Fe$_{1-x}$Co$_x$)$_2$As$_2$

Shyam Mohan[1], Yuji Tsuchiya[1], Yasuyuki Nakajima[1,2], and Tsuyoshi Tamegai[1,2]

[1]*Department of Applied Physics, University of Tokyo,7-3-1 Hongo, Bunkyo-ku, Tokyo 113-8656, Japan*

[2]*JST, Transformative Research-Project on Iron Pnictides (TRIP), 7-3-1 Hongo, Bunkyo-ku,Tokyo 113-8656, Japan*

We explore the flux front patterns during virgin penetration and after remagnetization in the iron-based superconductor Ba(Fe$_{1-x}$Co$_x$)$_2$As$_2$. After remagnetization we observe an instability characterized by turbulent dynamics at the vortex-antivortex boundary. Associated with the turbulent flux boundary a band of excess current density is observed. The turbulence is contrasted with the instability observed in the cuprate superconductors. Our results suggest turbulent instability at the vortex-antivortex interface may be a common feature in all remagnetized type-II superconductors.

The magnetic behavior of type-II superconductors is usually described in terms of the critical state model due to Bean[1] (which assumes the lower critical field $H_{c1}$ is zero and field independent critical current density $J_c$). In many different superconductors the critical state can break down under specific experimental conditions as is remarkably seen in the development of flux turbulence[2,3] and dendritic avalanches[4,5]. The flux turbulence is observed when the superconductor is remagnetized by a magnetic field of opposite polarity to the initially applied field. The sample then consists of vortices with opposite polarities (vortex and antivortex) separated by the remagnetization front. Ideally, in the absence of instabilities, it is expected that the flux penetration into the superconductor during remagnetization follows the same pattern as the virgin magnetization. However, in the 123-type cuprates (optimally doped and underdoped $YBa_2Cu_3O_7$ and $NdBa_2Cu_3O_7$) spectacularly different penetrations exhibiting turbulence, characterized by massive swirling of the flux front, have been observed upon remagnetization[2,6,7,8,9,10,11]. Vlasko-Vlasov et al.[7] argued that the remagnetization flux front in platelet-shaped superconductors leads to formation of a specific three-dimensional structure inside the sample. This structure called "Meissner hole" consists of flux-free regions around which closed vortex loops are formed that sustain a local increase of current. While the influence of the Meissner holes on the observed turbulence is widely accepted the actual mechanism of the instability is debated. Notably, Bass et al.[12,13] considers the instability to arise from overheating of the flux interface caused by vortex-antivortex annihilation and its relaxation. In contrast, Fisher et al.[8,9] analyzed this problem in terms of hydrodynamic turbulence at the boundary between two fluids due to a discontinuity of their tangential velocities; in anisotropic superconductors such a discontinuity is naturally favored. Additionally, mechanisms that consider a combination of both anisotropy and thermal effects have also been proposed[14,15].

Experimentally the fact that the turbulent instability is observed for crystals with different anisotropies and critical current densities suggests the intrinsic nature of such instabilities. Prompted by this possibility, we investigate the flux penetration in the widely studied iron-based superconductor $Ba(Fe_{1-x}Co_x)_2As_2$.

The iron-based superconductors have been subjected to intensive research due to their close comparison with the well-studied high-temperature cuprate superconductors[16]. The magnetic properties of electron-doped $Ba(Fe_{1-x}Co_x)_2As_2$ have been fairly well-investigated due to the ease of growing clean single crystals[17,18]. The main features of the vortex state in the slightly anisotropic ($\gamma \sim 3$) material $Ba(Fe_{1-x}Co_x)_2As_2$ and cuprate superconductors such as $YBa_2Cu_3O_7$ are similar: self-field $J_c \sim 10^6$ A/cm$^2$ at low temperatures, "fishtail effect" in magnetization hysteresis loops, evidence for predominantly collective vortex pinning, and giant flux creep in magnetic relaxation[18,19]. In this paper, we explore the flux penetration in single crystals of $Ba(Fe_{1-x}Co_x)_2As_2$, especially focusing on the influence of a remagnetized field on the nature of the flux front, using MO imaging. Hence our objective goes beyond the known features of the flux patterns in the mixed state previously obtained by MO imaging[18,19,20]. Our observations provide compelling evidence for the presence of turbulent flux instability in a superconductor other than the 123-type cuprates.

Single crystals of optimally doped $Ba(Fe_{0.925}Co_{0.075})_2As_2$ were grown using a self-flux method[18]. Bulk magnetization measurements revealed this batch had a critical temperature $T_c$ $\sim$24 K and self-field $J_c \sim 8\times10^5$ A/cm$^2$ at 5 K. Several single crystals from this batch were investigated. The results presented here pertain to a single crystal which was cleaved and cut to dimensions $\sim 530 \times 460 \times 30$ μm$^3$. MO imaging was performed in a flow-type helium cryostat using a Bi-doped ferrimagnet garnet film as the indicator. Using calibration methods, the two-dimensional pixel map of intensity was converted to the values of the local axial component of magnetic induction $B_z(x,y)$. All MO images were obtained with the external magnetic field applied parallel to the crystallographic *c* axis of the sample. MO images were captured with the polarizers aligned slightly away from the crossed position; in terms of image appearance this means that bright and dark correspond to positive and negative magnetic induction values, respectively.

Fig. 1(a) shows the remanent state in Ba(Fe$_{0.925}$Co$_{0.075}$)$_2$As$_2$ at 10 K captured after zero-field cooling followed by application and removal of +800 Oe field. When the field is removed, magnetic flux starts to exit from the sample leaving behind vortices trapped by the pinning sites. The image in Fig. 1(a) arises due to inability of the flux to fully penetrate the sample with 800 Oe field. The trapped flux patterns in Fig. 1(a) closely match that expected from the Bean critical state model with flux concentration at the sample center and diagonals. In particular, the penetration depth of the flux from all the edges is found to be nearly equal. The expelled flux concentrates at the sample edges and distinct irregular profiles of these antivortices can be seen in (a). This arises from preferential flux penetration along edge defects. Figs. 1(b) and (c) show the flux penetration into the single crystal in an applied field of +300 Oe at 10 K and 15 K, respectively. The irregular penetration seen at the edges of (a) is nearly sustained as it moves deeper into the sample. Remarkable distortions to the Bean-like state are clearly observed upon flux entry into the sample. The flux fronts are not smooth and have curvy features that are probably due to inhomogenous distribution of the strength / density of the pinning centers on the crystal surface. Fig 1(d) shows the line profiles obtained during virgin flux penetration at 15 K with increasing fields. At 100 Oe and 200 Oe the penetration profiles are as expected: smooth drops in the induction and progressive entry of flux to the sample interior with field. At 300 Oe the profile exhibits a plateau-like shape in regions where flux has already penetrated; the origin of this feature is unclear.

We now describe the observations that constitute the main focus of this paper, namely the influence of remagnetization in crystals of optimally-doped 122. Fig 2 (a)-(c) shows the MO images captured after remagnetization at 10 K, 15 K, and 17.5 K, respectively. The remagnetization is performed after cooling the sample in -300 Oe field to the measurement temperature. Subsequently, the magnetic field is reversed to +200 Oe and MO images are captured by averaging over several frames. The remagnetization flux front appears as the boundary between the dark and bright regions. The vortex-antivortex interface has become considerably wigglier and starts to curve around much more than the virgin flux front. The enhanced curvature and the distinct fingering of the interface are characteristic of a turbulent instability of the flux front upon remagnetization. Similar results were also obtained at lower temperatures down to 5 K. With increasing field/temperature the flux front loses some of its zig-zag curvy features instead developing into broader curves. Above 17.5 K the fingering flux patterns and the curvy nature of the flux front both gradually disappear from all edges; finally when the front reaches the sample center it gets progressively smoother. The flux front then resembles the virgin penetration at high temperatures/fields. On performing the remagnetization procedure on subsequent experimental runs different flux front profiles were observed. Fig 2(d) and (e) show the MO images after remagnetization with +100 Oe at 10 K and 15 K, respectively. These images were captured after zero-field-cooling to desired temperature, followed by application of -800 Oe field, and finally remagnetized with positive fields. It is evident that the flux front patterns across all edges are quite different than those observed in Fig 2(a)-(c).

Especially the nature of the flux fingering and the orientation of the fingers are in stark contrast with those in the left panels in Fig. 2.

Fig 3 (a)-(c) shows the magnetic induction profiles $B_z(x,y)$ across the dashed line in Fig 2(b) at 5 K, 10 K, and 15 K after cooling the crystal in -300 Oe field followed by application of different positive fields (same procedure used to obtain Fig 2(a)-(c)). Deep inside the sample the profiles correspond to the initially applied -300 Oe field where the reversed flux has still not entered. From the crystal edges the reversed flux progressively moves deeper inside the sample with increasing value of the reversed field. The most striking features in the $B_z(x,y)$ curves are the distinct steps in magnetic induction around $B_z = 0$, i.e. at the remagnetization flux boundaries. On either ends of the induction-steps small humps or upturns are clearly observed. These observations clearly point to the presence of excess currents around $B_z = 0$. Such a step at the flux boundary is usually explained as arising from excess surface-like currents due to finite $H_{c1}$ (the d$B$/d$H$ effect[21]); this can be understood by replacing the original Bean's critical state model with a triangular approximation[22]. At a given temperature the step-size has no observable dependence on the strength of the reversed field. With increasing temperature the flux front moves deeper into the sample while sustaining the distinct magnetic induction step at the boundary. The magnitude of the induction-step drops with increasing temperature, from ~400 G at 10 K to ~180 G at 17.5 K. Finally the induction steps vanish when the reversed flux fully envelops the sample. To obtain a qualitative picture of the excess current flow associated with the observed local magnetic induction steps we show maps of the in-plane critical current density $J_{xy} = (J_x^2 + J_y^2)^{1/2}$ in Fig 2(f) at 15 K. $J_{xy}$ is obtained from the corresponding magnetic induction map (Fig. 2(b)) by inversion of the Biot-Savart's law assuming two-dimensional current flow[23]. Here the brightness of the grayscale images is proportional to the magnitude of the current density. At the location of the vortex-antivortex boundaries a band of excess current is vividly observed. This confirms that the instability, observed upon remagnetization, is associated with the appearance of "Meissner holes" at the flux front around which excess surface-like currents flow.

We now examine the similarities/differences in the nature of flux penetration instabilities upon remagnetization in Ba(Fe$_{0.925}$Co$_{0.075}$)$_2$As$_2$ and the cuprate superconductors. (i) At the temperatures studied here the critical current density of the crystal is ~$10^5$ A/cm$^2$ which is of similar range to that in YBa$_2$Cu$_3$O$_7$ in the temperature region where turbulence is observed. (ii) The anisotropy in Ba(Fe$_{0.925}$Co$_{0.075}$)$_2$As$_2$ is nearly half of that in the 123-cuprates which suggests that the instability is not coupled with high anisotropy. (iii) The skeleton of the curvy /wiggly nature is found to exist even in the virgin flux penetration images in Ba(Fe$_{0.925}$Co$_{0.075}$)$_2$As$_2$. Even then, upon remagnetization the flux turbulence is easily discernible by the enhanced bending of the flux front. Thus the remagnetization in Ba(Fe$_{0.925}$Co$_{0.075}$)$_2$As$_2$ is significantly influenced by the defect density and strength. (iv) While twin boundaries are ubiquitous in 123-type cuprates, stripes associated with a structural transition in 122 above $T_c$ are observed only in underdoped samples[24]. However, this strongly correlates with the observation of flux turbulence in twinless YBa$_2$Cu$_3$O$_7$ single crystals emphasizing that any defects can lead to flux-front distortion[25]. (v) A characteristic signature in the cuprate samples exhibiting flux turbulence concerns the time evolution of the vortex-antivortex boundary[10,11,23]. To study the time evolution in Ba(Fe$_{0.925}$Co$_{0.075}$)$_2$As$_2$ the sample is remagnetized as described earlier (see discussion of Figs 2 (a)-(c)). Following the application field-reversal a sequence of images were captured for times up to 60 s as the flux front moves deeper into the sample center. Fig. 4(b) shows the evolution of the magnetic induction profiles at 15 K after applying a reversed field of +300 Oe. For comparison, shown in Fig. 4(a) is the evolution of the profiles after virgin magnetization at +300 Oe. In both cases, significant flux front motion is completed within ~20 s as seen by the large spacing between the profiles in Fig. 4. Thereafter the front motion proceeds in a steady-state slower manner. Thus the time-development of the flux fronts proceed in identical fashion irrespective of the field history. This observation suggests that pinning in Ba(Fe$_{0.925}$Co$_{0.075}$)$_2$As$_2$ is strong enough to suppress any anomalous flux motion associated with the instability. (vi) Lastly, flux turbulence in the cuprates has always been observed only in specific temperature windows. In our case, at temperatures above 20 K the features are blurred out due to both the increased flux motion and the relative loss in contrast of the image. Thus it appears that there is an upper temperature limit for the appearance of instability. On the other hand, we notice significant turbulence that persists down to the lowest measurement temperatures (5 K).

To summarize, we observed instabilities in the vortex state of the iron-based superconductor Ba(Fe$_{0.925}$Co$_{0.075}$)$_2$As$_2$. The instability manifests as turbulent dynamics of the vortex-antivortex boundary in single crystals that are subjected to remagnetization. The turbulent flux penetration patterns are different from the virgin penetration and do not repeat themselves. Associated with the turbulence we also find at the boundary the existence of steps in magnetic induction which arise from excess current density at the flux front. This excess band of currents could significantly affect the magnetic relaxation of vortices leading to the observed low-field anomalous features in Ba(Fe$_{0.925}$Co$_{0.075}$)$_2$As$_2$ crystals[26,27] and FeTe$_x$Se$_{1-x}$[28]. Our results provide new vistas of a strong similarity between the vortex state properties in the iron-based superconductors with the high-temperature cuprate superconductors. We surmise that turbulent instability upon remagnetization is not confined to the cuprate superconductors but is a basic feature of all type-II superconductors.

SM gratefully acknowledges the Japan Society for the Promotion of Science for support.

**References:**

[1] C. P. Bean, Rev. Mod. Phys. **36**, 31 (1964).

[2] V. K. Vlasko-Vlasov, V. I. Nikitenko, A. A. Polyanskii, G. Crabtree, U. Welp, and B.W. Veal, Physica C **222**, 361 (1994).

[3] M.V. Indenbom, Th. Schuster, M. R. Koblischka, A. Forkl, H. Kronmuller, L. A. Dorosinskii, V. K. Vlasko-Vlasov, A. A. Polyanskii, R. L. Prozorov, and V. I. Nikitenko, Physica C **209**, 259 (1993).

[4] C. A. Duran, P. L. Gammel, R. E. Miller, and D. J. Bishop, Phys. Rev. B **52**, 75 (1995).

[5] T. H. Johansen, M. Baziljevich, D. V. Shantsev, P. E. Goa, Y. M. Galperin, W. N. Kang, H. J. Kim, E. M. Choi, M. S. Kim, and S. I. Lee, Europhys. Lett. **59** (4), 599 (2002).

[6] V. K. Vlasko-Vlasov, U. Welp, G. W. Crabtree, D. Gunter, V. Kabanov and V. I. Nikitenko Phys. Rev. B **56**, 5622 (1997).

[7] V. K. Vlasko-Vlasov, U.Welp, G.W. Crabtree, D. Gunter, V.V. Kabanov, V. I. Nikitenko, and L.M. Paulius, Phys. Rev. B **58**, 3446 (1998).

[8] L. M. Fisher, P. E. Goa, M. Baziljevich, T. H. Johansen, A. L. Rakhmanov, and V. A. Yampolskii, Phys. Rev. Lett. **87**, 247005 (2001).

[9] L. M. Fisher, A. Bobyl, T. H. Johansen, A. L. Rakhmanov, V. A. Yampolskii, A. V. Bondarenko, and M. A. Obolenskii, Phys. Rev. Lett. **92**, 247005 (2001).

[10] T. Frello, M. Baziljevich, T. H. Johansen, N. H. Andersen, T. Wolf, and M. R. Koblischka Phys. Rev. B **59**, R6639 (1999).

[11] I. F. Voloshin, A. V. Kalinov, L. M. Fisher, V. A. Yampolskii, A. Bobyl, and T. H. Johansen, Low Temp. Phys. **35**, 627 (2009).

[12] F. Bass, B. Ya. Shapiro, and M. Shvartser, Phys. Rev. Lett. **80**, 2441 (1998).

[13] F. Bass, B. Ya. Shapiro, I. Shapiro, and M. Shvartser, Phys. Rev. B **58**, 2878 (1998).

[14] C. Baggio, M. Howard, and W. van Sarloos, Phys. Rev. E **70**, 026209 (2004).

[15] E. E. Dvash, I. Shapiro, B. Rosenstein, and B. Ya. Shapiro, arXiv:1109.0637

[16] I. I. Mazin, Nature **464**, 183 (2010).

[17] A. S. Sefat, R. Jin, M. A. McGuire, B. C. Sales, D. J. Singh, and D. Mandrus, Phys. Rev. Lett. **101**, 117004 (2008).

[18] Y. Nakajima, T. Taen, and T. Tamegai, J. Phys. Soc. Jpn. **78**, 023702 (2009).

[19] R. Prozorov, N. Ni, M. A. Tanatar, V. G. Kogan, R. T. Gordon, C. Martin, E. C. Blomberg, R. Prommapan, J. Q. Yan, S. L. Bud'ko, A. I. Goldman, and P. C. Canfield, Phys. Rev. B **78**, 224506 (2008).

[20] Z. W. Lin, J. Zhu, Y. Guo, Y. Li, S. Wang, Y. B. Zhang, K. X. Xu, and C. B. Cai, J. Appl. Phys. **107**, 09E155 (2010).

[21] M. A. R. LeBlanc, S. X. Wang, D. LeBlanc, M. Krzywinski, and J. Meng, Phys. Rev. B **52**, 12895 (1995).

[22] M.V. Indenbom, Th. Schuster, H. Kuhn, H. Kronmuller, T. W. Li, and A. A. Menovsky, Phys. Rev. B **51**, 15484 (1995).

[23] R. J. Wijngaarden, H. J. W. Spoelder, R. Surdeanu, and R. Griessen, Phys. Rev. B **54**, 6742 (1996).

[24] R. Prozorov, M. A. Tanatar, N. Ni, A. Kreyssig, S. Nandi, S. L. Bud'ko, A. I. Goldman, and P. C. Canfield, Phys. Rev. B **80**, 174517 (2009).

[25] L. S. Uspenskaya, I. G. Naumenko, and A. A. Zhokov, Physica C **402**, 188 (2004).

[26] Y. Nakajima, Y. Tsuchiya, T. Taen, T. Tamegai, S. Okayasu, and M. Sasase, Phys. Rev. B **80**, 012510 (2009).

[27] T. Tamegai, T. Taen, H. Yagyuda, Y. Nakajima, Y. Okayasu, M. Sasase, H. Kitamura, T. Murakami, T. Kambara, and T. Kanai, Physica C **402**, doi:10.1016/j.physc.2011.05.052.

[28] T. Taen, Y. Nakajima, and T. Tamegai, Physica C **470**, 1106 (2010).

**Figure Captions:**

**Figure 1:** (a) MO image of the remanent state in Ba(Fe$_{0.925}$Co$_{0.075}$)$_2$As$_2$ crystal at 10 K. Virgin flux penetration images obtained after zero-field cooling followed by application of 300 Oe field at (b) 10 K, and (c) 15K. (d) Magnetic induction profiles during virgin flux penetration at 15 K across the dashed line in (c) for different applied fields. The dashed lines in (d) denote the sample edges.

**Figure 2:** (a)-(c) MO images obtained after remagnetization in a Ba(Fe$_{0.925}$Co$_{0.075}$)$_2$As$_2$ crystal at 10 K, 15 K, and 17.5 K, respectively, after field cooling in -300 Oe and subsequent field reversal to 200 Oe. Remagnetization images captured after zero-field cooling, followed by application of -800 Oe field and subsequent reversal to 100 Oe at (d) 10 K, and (e) 15 K, respectively. (f) The calculated current density after inverting the magnetic induction map of Fig. 2(b).

**Figure 3:** Local magnetic induction profiles after remagnetization across the dashed line in Fig. 2(b) at (a) 10 K, (b) 15K, and (c) 17.5 K after cooling in -300 Oe field followed by reversing the field polarity from 0 Oe to 300 Oe in steps of 50 Oe. Dashed lines denote sample edges.

**Figure 4:** The time evolution of the induction profiles at 15 K with 300 Oe field applied (a) during virgin penetration, and (b) after remagnetization, respectively, from 0 s to 62 s. Dashed lines denote sample edges.

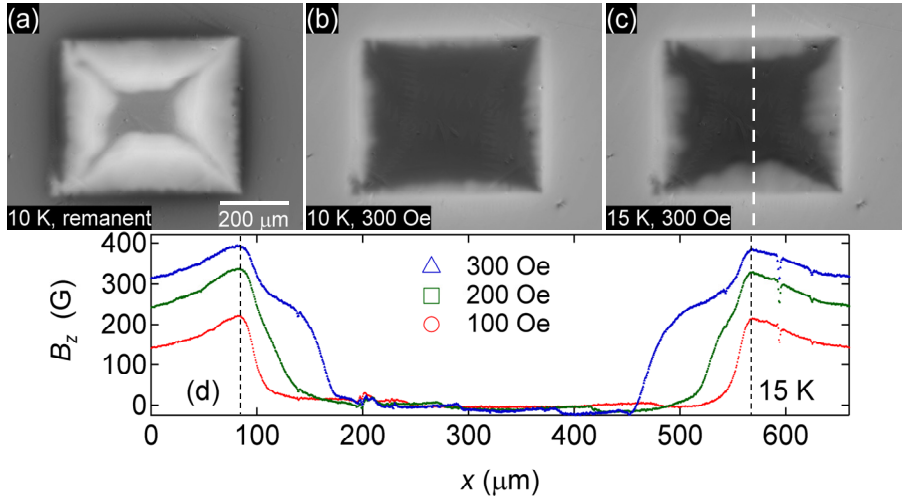

Figure 1

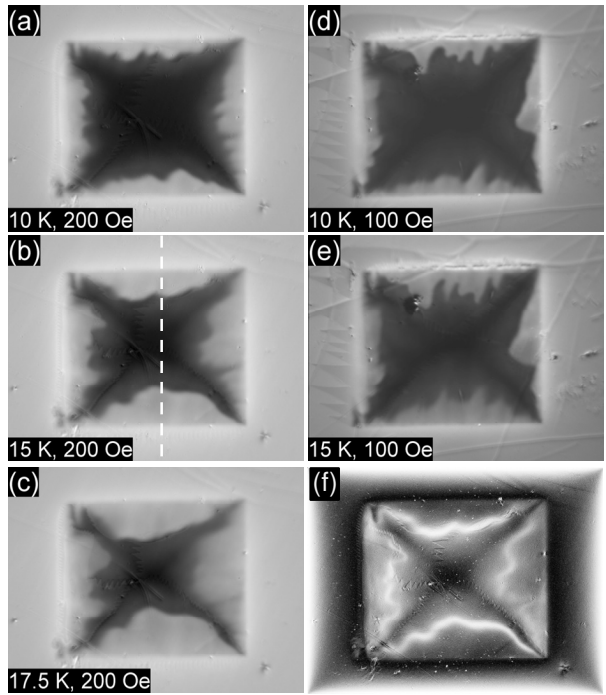

Figure 2

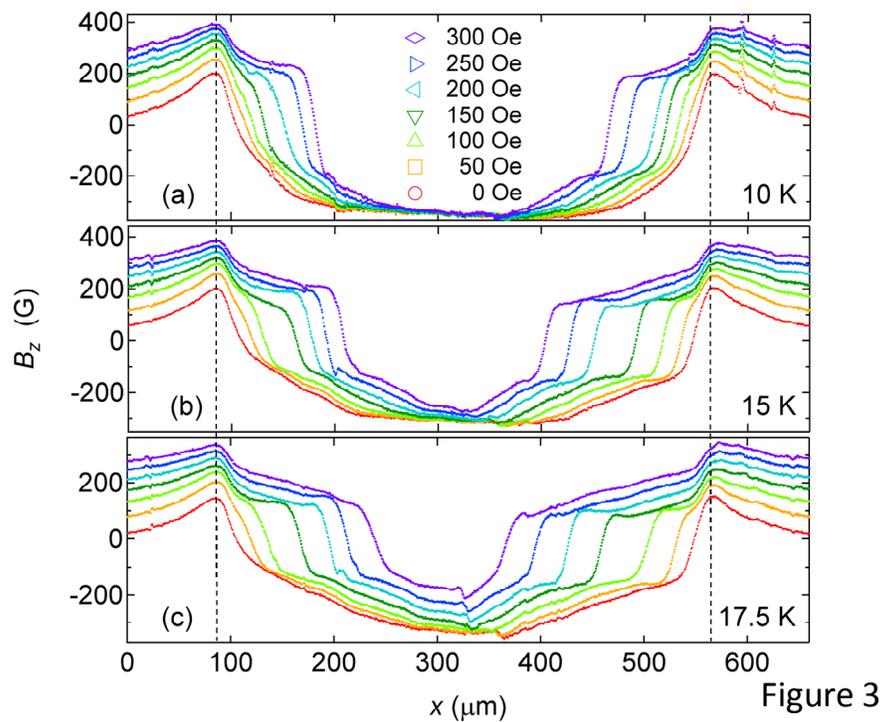

Figure 3

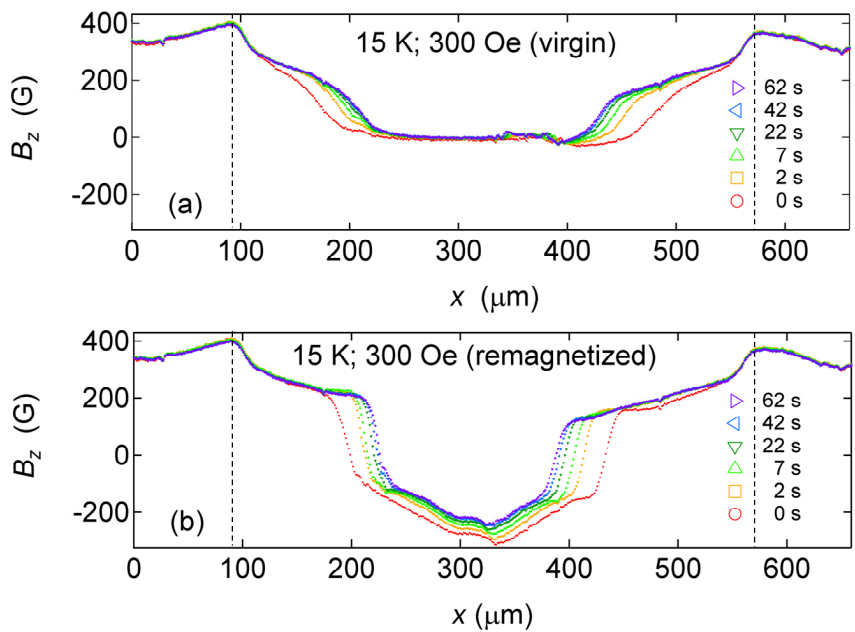

Figure 4